\documentclass[aps,prd,reprint,superscriptaddress,nofootinbib]{revtex4-1}
\pdfoutput=1
\usepackage[T1]{fontenc}
\usepackage{setspace}
\usepackage{amsmath}
\usepackage{graphicx}
\usepackage{placeins}
\usepackage{units}
\usepackage{xspace}
\newcommand{\nue}{\ensuremath{\nu_e}\xspace}
\newcommand{\numu}{\ensuremath{\nu_\mu}\xspace}
\newcommand{\ccpi}{\ensuremath{\nu_{\mu}p\rightarrow \mu^{-}p\pi^{+}}\xspace}
\usepackage{tabulary}
\usepackage{caption}
\usepackage{subcaption}
\usepackage{multirow}
\usepackage{color} 
\begin{document}

\title{Reanalysis of bubble chamber measurements of muon-neutrino induced single pion production}
\date{\today}
\author{Callum Wilkinson}
\affiliation{University of Sheffield}
\author{Philip Rodrigues}
\affiliation{University of Rochester}
\author{Susan Cartwright}
\affiliation{University of Sheffield}
\author{Lee Thompson}
\affiliation{University of Sheffield}
\author{Kevin McFarland}
\affiliation{University of Rochester}
\begin{abstract}
There exists a longstanding disagreement between bubble chamber
measurements of the single pion production channel \ccpi from the
Argonne and Brookhaven National Laboratories. We digitize and
reanalyse data from both experiments to produce cross-section ratios
for various interaction channels, for which the flux uncertainties
cancel, and find good agreement between the experiments. By multiplying the cross-section
ratio by the well-understood charged current quasi-elastic
cross-section on free nucleons, we extract single-pion production
cross-sections which do not depend on the flux normalization
predictions. The \ccpi cross-sections we extract show good agreement
between the ANL and BNL datasets.
\end{abstract}
\maketitle

\section{Introduction}
Single pion production by neutrinos is an important process at neutrino
energies around \unit[1]{GeV}, where the dominant production mechanism
is via the production and subsequent decay of hadronic resonances. In
neutrino oscillation experiments, neutral-current neutral pion production is a
background to \nue charged-current events in $\numu \to \nue$
measurements, while charged-current events producing charged pions
contribute to $\numu$ disappearance measurements, either as background
in analyses which select quasi-elastic events, or as signal in
analyses which use an inclusive charged-current selection.

Predictions of single pion production on the nuclei used in neutrino
oscillation experiments usually factorize the modelling into three parts: the neutrino-nucleon cross section;
additional effects due to the nucleon being bound in the nucleus; and
the ``final state interactions'' (FSI) of hadrons exiting the
nucleus. Experimental knowledge of the neutrino-nucleon cross section
for single pion production, in the $\unit[100]{MeV}$ to few-\unit{GeV}
neutrino energy range relevant for current and planned oscillation
experiments, is sparse, coming from bubble chamber experiments with
hydrogen or deuterium targets with low statistics. In particular, data
from the \unit[12]{ft} bubble chamber at Argonne National Laboratory
(ANL) and the \unit[7]{ft} bubble chamber at Brookhaven National
Laboratory (BNL) for the leading single pion production process \ccpi differ in normalization by
\unit[30--40]{\%}. This data is used to constrain the axial form factor for pion production on 
free nucleons, which cannot be constrained by electron scattering data, so this discrepancy leads 
to large uncertainties in the predictions for oscillation experiments
~\cite{sobczyk_2009, hernandez_spp_2007, leitner_spp_2008, hernandez_spp_2010, k2k_2003, t2k-nue-prd}, 
as well as in interpretation of data taken on nuclear targets~\cite{gibuu_spp_tune}.

Resolving this discrepancy will be vital for current and future
neutrino oscillation experiments, which have very stringent systematic
error requirements~\cite{lbne,t2k_sensitivity_2014}, but current
neutrino cross section measurements are taken on nuclear targets such
as carbon and oxygen, where it is difficult to disentangle the
neutrino-nucleus cross section from the effects of the nucleus and
FSI. In this context, it is worthwhile to revisit the ANL and BNL
datasets to look for possible consistency. Graczyk \emph{et~al.}\ have
found consistency in the datasets by carefully considering
normalization uncertainties~\cite{sobczyk_2009}\cite{sobczyk_2014} and
deuteron nuclear effects. In this paper, we present a complementary
approach in which we consider ratios of event rates for different
processes in the ANL and BNL experiments, in which normalization
uncertainties cancel. By multiplying the event rate ratio by an
independent measurement of the cross section of the denominator, we
obtain a measurement of the single-pion production cross section. In
essence, this method amounts to using the denominator cross section as
the factor which converts an event rate into a cross section, where
the original analyses used a prediction of the neutrino flux for the
same purpose.

The paper is organized as follows. In Section~\ref{sec:data}, we describe the method for obtaining the data from the original papers. A discussion of the sources of error for these datasets is in Section~\ref{sec:error}. In Section~\ref{sec:ratios}, we present the ratios of event rates for various processes. Then these are used in Section~\ref{sec:cross_sections} to extract CC-inclusive and \ccpi cross-sections, where we find good agreement between ANL and BNL. Our conclusions are presented in Section~\ref{sec:conclusions}.

\section{Obtaining data}\label{sec:data}
A literature review of the ANL and BNL cross-section papers produces a wealth of data. For this analysis, corrected event rates as a function of the neutrino energy, $E_{\nu}$, are required. Corrected event rates are obtained from the raw (measured) event rates by estimating detector inefficiencies, and subtracting background processes. As we are interested in $\nu_{\mu}-\mathrm{D_{2}}$ interactions, we remove data from hydrogen fills of the experiments. This section discusses how the datasets used in this analysis were obtained. All data published as histograms have been digitized using the engauge digitizer tool~\cite{engauge}. The discrepancy between the published event rates and the event rate obtained by integrating the digitized histograms is less than 1\%. The effect of ditigization on the shape of the distributions is assumed to be small.

\subsection{ANL \unit[12]{ft} bubble chamber}
A description of the experimental setup for the \unit[12]{ft} bubble chamber at Argonne National Laboratory (ANL) can be found in~\cite{ANL_Barish_1977}. Additional details of the event reconstruction and classification algorithms used can be found in~\cite{ANL_Radecky_1982} and~\cite{ANL_Barish_1979}. In the ANL experiment, data was initially taken with a hydrogen fill of the bubble chamber, then data was taken with a deuterium fill for the remainder of the experiment~\cite{ANL_Barish_1977}. Event rates are only available as a combination of both hydrogen and deuterium fills of the detector, so care must be taken to remove the hydrogen component. Published cross-sections are given using two different datasets, which we refer to here as the ANL partial and full datasets. The partial dataset is described in Reference~\cite{ANL_Barish_1977}, and is approximately 30\% of the final dataset. The full dataset is described in Reference~\cite{ANL_Radecky_1982}. Events on hydrogen comprise approximately 2\% (6\%) of the total for the full (partial) dataset.

The raw event rate for the \ccpi channel is given in~\cite{ANL_Radecky_1982} using the complete ANL dataset. No invariant mass cuts were used when selecting these events. The published (digitized) number of events before corrections is 871 (843.2); we scale the digitized distribution to the published corrected event rate of 1115.0. A small subset of the data comes from the earlier hydrogen fill of the detector, which contributes 90 \ccpi events (corrected)~\cite{ANL_Barish_1979}.

The corrected event rates for the CCQE and CC-inclusive channels with the partial ANL datset are taken from~\cite{ANL_Barish_1979}. The events are presented as four samples, as summarized in Table~\ref{tab:ANL_log_validation}: in later stages of the analysis, the digitized distributions are scaled to match the published event rate. The CC-inclusive contribution from proton interactions is also scaled to remove the 102 interactions on hydrogen (of 457 total).

\begin{table}[h]
  \centering
  {\renewcommand{\arraystretch}{1.2}
    \begin{tabular}{|c|c|c|c|}
      \hline
      Dataset & Channel & Digitized & Published \\
      \hline
      \multirow{2}{*}{Partial} & $\nu n \rightarrow \mu^{-}p$  & 834.6     & 833       \\
      & $\nu p \rightarrow \mu^{-}p\pi^{+}$ & 395.9     & 398       \\
      & $\nu n \rightarrow \mu^{-} X^{+}$   & 1139.2    & 1150      \\
      & $\nu p \rightarrow \mu^{-} X^{++}$  & 453.2     & 457       \\
      \hline
      Full & $\nu p \rightarrow \mu^{-}p\pi^{+}$ & 843.2 & 871 \\
      \hline
  \end{tabular}}
  \caption{Numbers of events for each of the ANL samples as published by ANL and as digitized for this work.}
  \label{tab:ANL_log_validation}
\end{table}

As the \ccpi event rate is given for both the partial and final ANL datasets, the ratio can be used to scale the CCQE and CC-inclusive samples from the partial dataset to the statistics of the full ANL dataset. The digitized ANL data for all channels considered, with all corrections applied, are given in Figure~\ref{subfig:ANL_rates}, where the errors are statistical only.

\subsection{BNL \unit[7]{ft} bubble chamber}
\label{sec:bnl}
A description of the experimental setup of the \unit[7]{ft} bubble chamber at Brookhaven National Laboratory (BNL), and a description of the event reconstruction and classification algorithms, can be found in References~\cite{BNL_Baker_1981,BNL_Baker_1982,BNL_Kitagaki_1990,BNL_Kitagaki_1986}. Although events were initially taken with a hydrogen fill of the bubble chamber, most BNL results are separated into hydrogen and deuterium measurements. As for ANL, published BNL cross-sections are given using two different datasets, which we refer to here as the BNL partial and full datasets. The partial dataset is described in Reference~\cite{BNL_Baker_1981}, and is approximately 30\% of the final dataset. The full dataset is described in Reference~\cite{BNL_Kitagaki_1986}.

The published (digitized) number of uncorrected CCQE events on deuterium, using the full BNL dataset, is 2684 (2693.3)~\cite{BNL_Kitagaki_1990}; we scale the digitized distribution by the published correction factor of 1.11 to obtain the corrected event rate. 

For \ccpi with no cut on the invariant mass, the published (digitized\footnote{The data was only extracted for $0 \leq E_{\nu} \leq 6$~\unit{GeV}, as it was difficult to distinguish higher energy bins reliably due to the quality of the published histogram, and so does not contain all of the events in the quoted raw event number. For this reason, the digitized histogram was not scaled to match the quoted event rate.}) number of raw events, for the complete BNL dataset, is 1610 (1534.7)~\cite{BNL_Kitagaki_1986} and the digitized distribution is scaled by the published correction factor of 1.123 to obtain the corrected event rate. The \ccpi results are also presented in Reference~\cite{BNL_Kitagaki_1990} as well as a \ccpi$/\mathrm{CCQE}$ ratio, but these results have a hadronic invariant mass cut, $W < \unit[1.4]{GeV}$, so have not been used for this analysis.

The uncorrected CC-inclusive event rate for $E_{\nu} \leq \unit[14]{GeV}$ using approximately 30\% of the total deuterium data is 3723 published~\cite{BNL_Baker_1982} and 3685.3 digitized. The correction factor for BNL CC-inclusive events is not explicitly given, but we can identify three corrections that should be applied to CC-inclusive data:
\begin{itemize}
  \item Scanning-measuring efficiency, $f_{1} = 1.11\pm0.02$~\cite{BNL_Kitagaki_1986},
  \item NC Background, $f_{2} = 0.94\pm0.01$~\cite{BNL_Baker_1981},
  \item H$_2$ contamination in D$_2$, $f_{3} = 0.96$.\footnote{The correction factor for H$_2$ contamination in D$_2$ is given as $0.87\pm0.02$ in~\cite{BNL_Kitagaki_1986}, for interactions off a proton. In~\cite{BNL_Baker_1982}, BNL measure the ratio of CC-inclusive reactions off a neutron to those off a proton as $\sigma(\nu n)/\sigma(\nu p) = 1.95 \pm 0.10$. We combine these to arrive at an estimate of the correction factor for H$_2$ contamination in D$_2$ for CC-inclusive events, $f_{3} = 0.96$.}
\end{itemize}
These corrections are combined to give a total correction factor $f \equiv f_1\times f_2 \times f_3 = 1.00$ for the CC-inclusive dataset.

\begin{table}[h]
  \centering
  {\renewcommand{\arraystretch}{1.2}
    \begin{tabular}{|c|c|c|c|}
      \hline
      Dataset & Channel  & Digitized & Published \\
      \hline
      \multirow{2}{*}{Partial} & $\nu n \rightarrow \mu^{-}p$  & ---  & 1276  \\
      & $\nu N \rightarrow \mu^{-} X$ & 3685.3  & 3723    \\
      \hline
      \multirow{2}{*}{Full} & $\nu n \rightarrow \mu^{-}p$        & 2693.3     & 2684       \\
      & $\nu p \rightarrow \mu^{-}p\pi^{+}$ & 1534.7 & 1610 \\
      \hline
  \end{tabular}}
  \caption{Numbers of events for each of the BNL samples as published by BNL and as digitized for this work. Note that it was not necessary to digitize the CCQE event rate for the partial BNL dataset.}
  \label{tab:BNL_event_rates}
\end{table}

\begin{figure}[htbp]
  \centering
  \centering
  \begin{subfigure}{0.45\textwidth}
    \includegraphics[width=\textwidth]{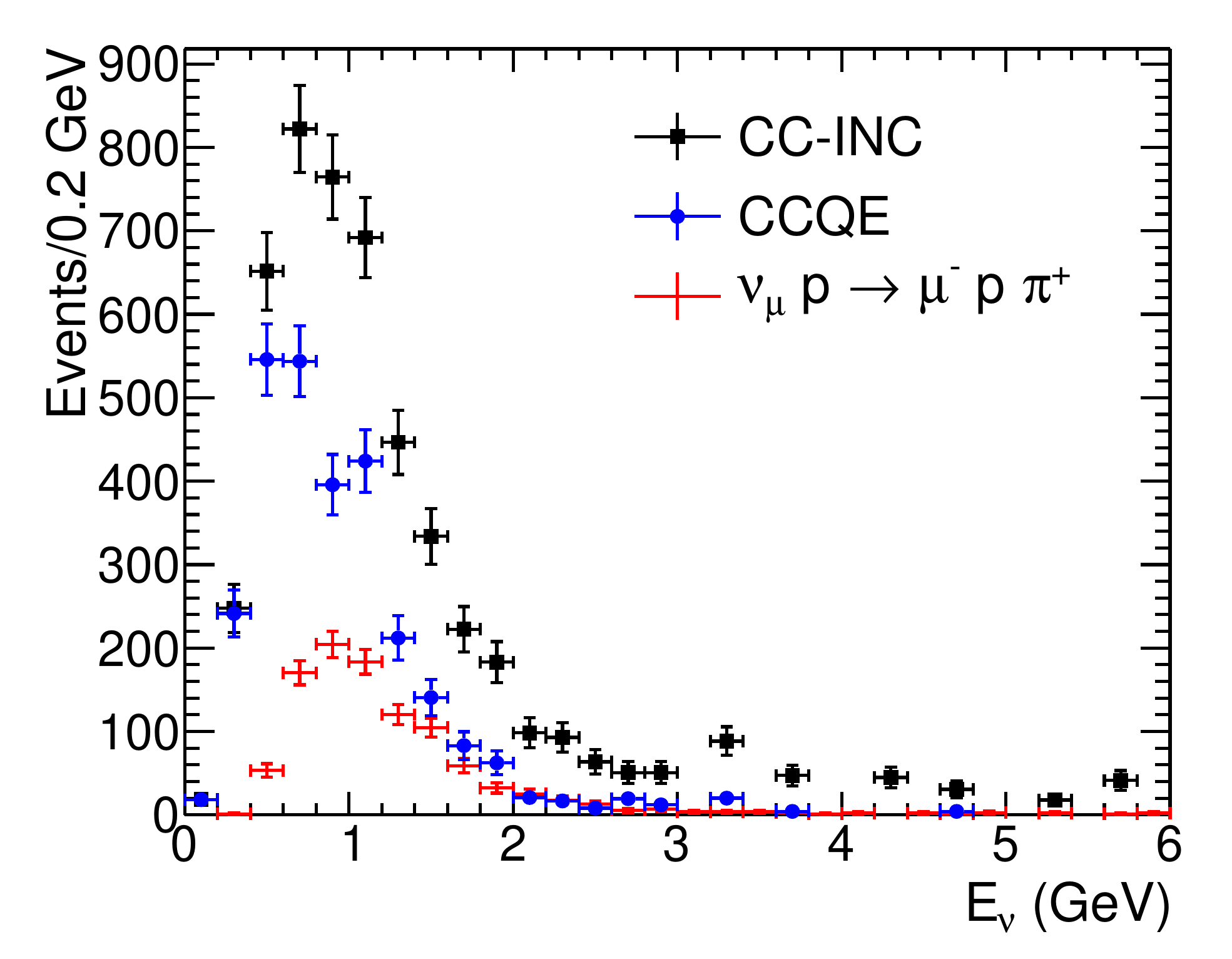}
    \caption{ANL}
    \label{subfig:ANL_rates}
    \end{subfigure}
  \begin{subfigure}{0.45\textwidth}
    \includegraphics[width=\textwidth]{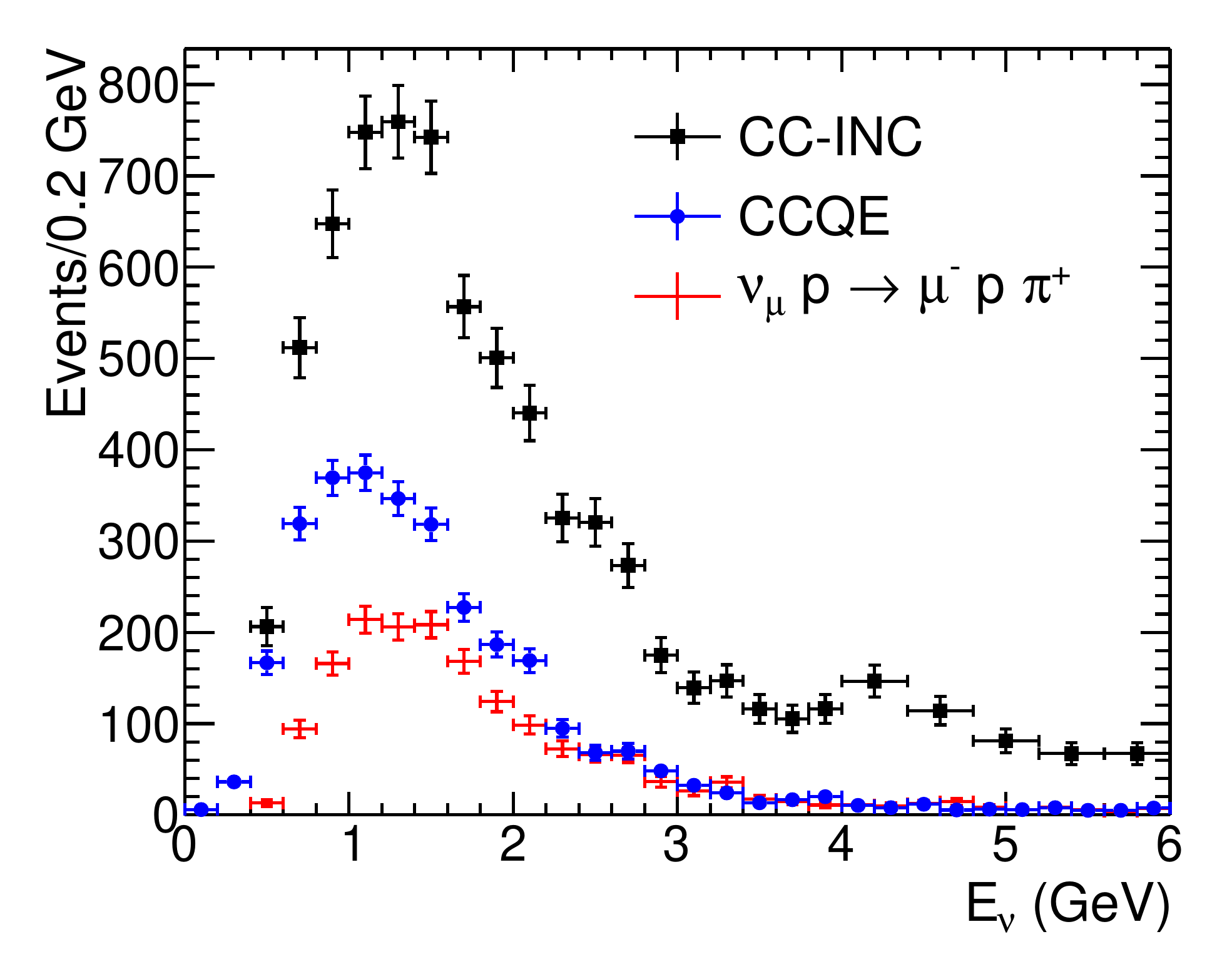}
    \caption{BNL}
    \label{subfig:BNL_rates}
    \end{subfigure}
  \caption{The digitized event rates on deuterium for the three interaction channels CCQE, \ccpi and CC-inclusive, as a function of the reconstructed neutrino energy $E_{\nu}$. The errors are statistical only. Both ANL and BNL event rates and errors have been scaled when necessary to the statistics of their full deuterium samples.}\label{fig:event_rates}
\end{figure}

The raw event rates are summarized in Table~\ref{tab:BNL_event_rates} for all processes. As the number of CCQE events is given for both the partial and complete datasets, the ratio of the two can be used to scale the CC-inclusive event rate up to the statistics of the full BNL deuterium dataset. The digitized BNL data for all channels, with all corrections applied, are given in Figure~\ref{subfig:BNL_rates}, shown with statistical errors only.

\section{Error analysis}\label{sec:error}
Throughout this work, only statistical errors are considered, which are the dominant source of error for these low-statistics bubble chamber datasets. Flux normalization errors are the second largest errors in the original ANL and BNL analyses, at around 15-20\%. These are not considered here because they cancel by construction in ratios of event rates.

There are additional errors on the overall normalization of all channels, which are introduced by the background subtraction and correction for detector effects; these are summarized for BNL in~\cite{BNL_Kitagaki_1990}, and can be inferred for ANL from~\cite{ANL_Barish_1979}. A conservative estimate of the normalisation error for both experiments would be approximately \unit[5]{\%}. It is also likely that many of the sources of uncertainty are common between interaction channels, and would cancel in the ratios calculated here, but a full error analysis is not possible.

There is also an error on the reconstructed neutrino energy, which is estimated for BNL to be $\frac{\Delta E_{\nu}}{E_{\nu}}\sim$\unit[2]{\%} for CCQE and \ccpi events~\cite{BNL_Baker_1981}, and $\sim$\unit[5]{\%} for other charged current production channels which are not kinematically overconstrained. ANL also quote an error of $\frac{\Delta E_{\nu}}{E_{\nu}}\leq$ \unit[5]{\%} for the harder to reconstruct channels, but do not quote an error on kinematically overconstrained channels~\cite{ANL_Radecky_1982}. As the uncertainty on $E_{\nu}$ largely comes from uncertainty in the beam direction, and BNL (ANL) quote small uncertainties $\pm 0.5^{\circ}$~\cite{BNL_Baker_1981} ($\pm 1.0^{\circ}$~\cite{ANL_Barish_1979}), we conclude that this error will be small, and $\frac{\Delta E_{\nu}}{E_{\nu}} \leq$ \unit[5]{\%} for all channels considered here.

\FloatBarrier
\section{$E_{\nu}$-dependent event rate ratios}\label{sec:ratios}
The number of events $N_X(E)$ for a given process $X$ in an energy bin $E$ is the product of flux $\Phi(E)$ and cross section $\sigma_X(E)$, so the ratio of corrected event rates for different channels is equal to the ratio of the cross-section for those channels. More importantly, in this ratio the flux and associated flux uncertainties cancel. So by taking the ratios between channels, it is possible to look for consistency in the ANL and BNL results regardless of possible problems with their flux predictions.
\begin{figure}[htbp]
  \centering
  \includegraphics[width=0.45\textwidth]{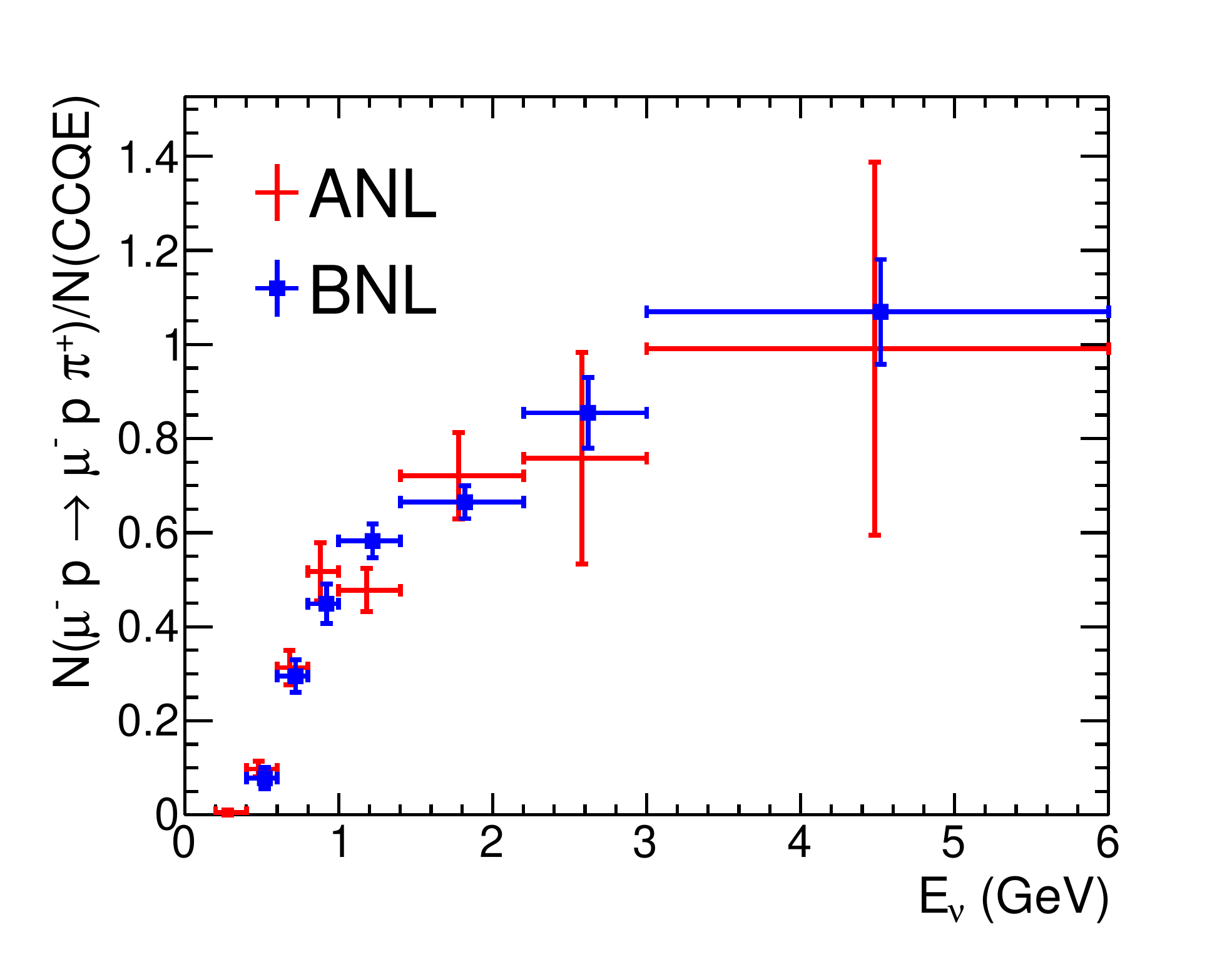}
  \caption{Ratio of \ccpi to CCQE events as a function of $E_{\nu}$ for both ANL and BNL.}\label{fig:CC1pi_CCQE_ratio}
\end{figure}

There is very good agreement between ANL and BNL in the ratio $\frac{\mathrm{CC}1\pi^{+}}{\mathrm{CCQE}}$, as shown in Figure~\ref{fig:CC1pi_CCQE_ratio}. This is contrary to expectation, given the discrepancy in the published \ccpi results, and suggests that the cause of the discrepancy is the flux prediction used to extract cross-sections from each experiment. This conclusion is supported by other analyses~\cite{sobczyk_2009}\cite{sobczyk_2014}, which found that the ANL and BNL results are compatible within their flux normalization uncertainties.
\begin{figure}[htbp]
  \centering
  \includegraphics[width=0.45\textwidth]{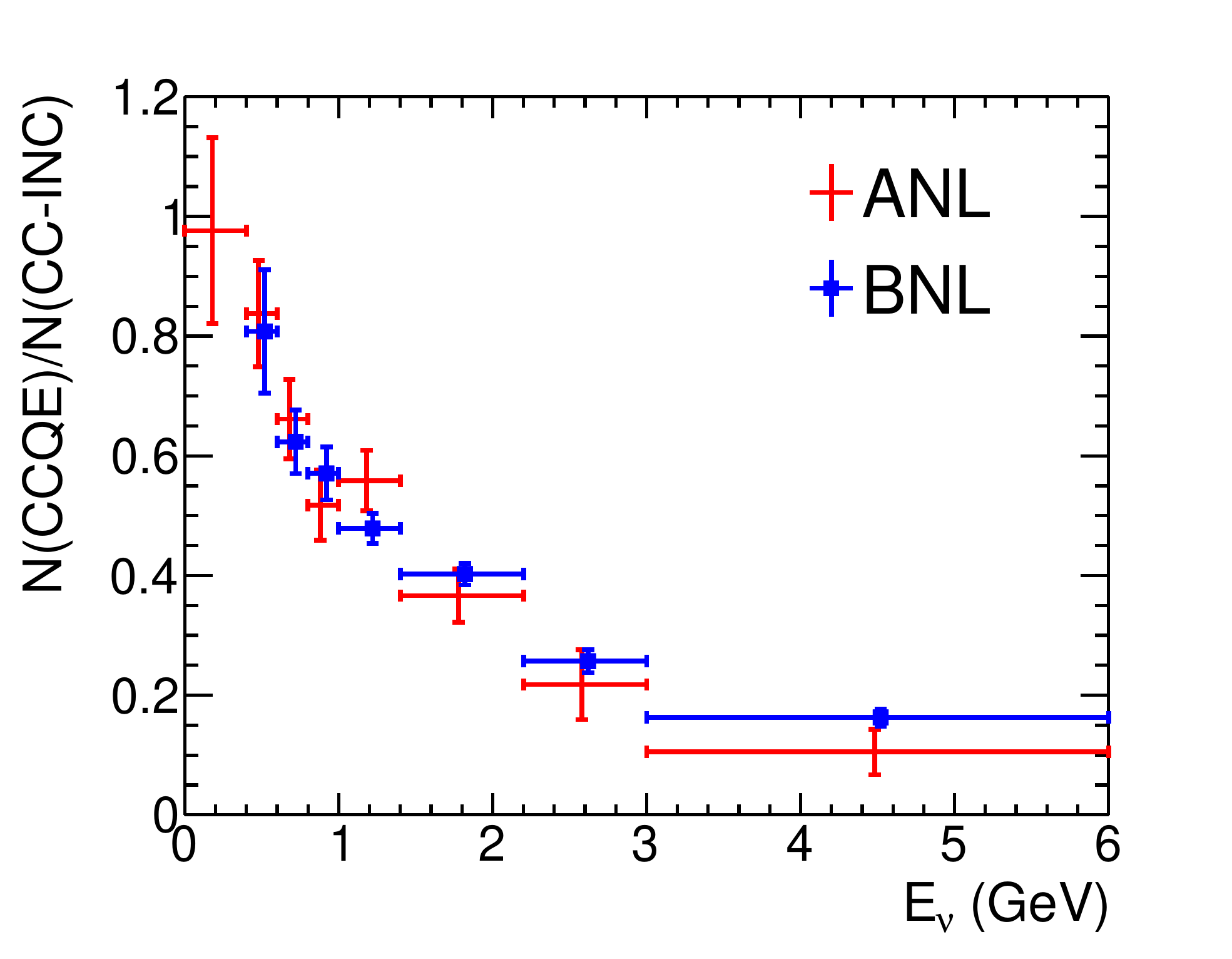}
  \caption{Ratio of CCQE to CC-inclusive events as a function of $E_{\nu}$ for both ANL and BNL.}\label{fig:CCQE_CCINC_ratio}
\end{figure}

\begin{figure}[htbp]
  \centering
  \includegraphics[width=0.45\textwidth]{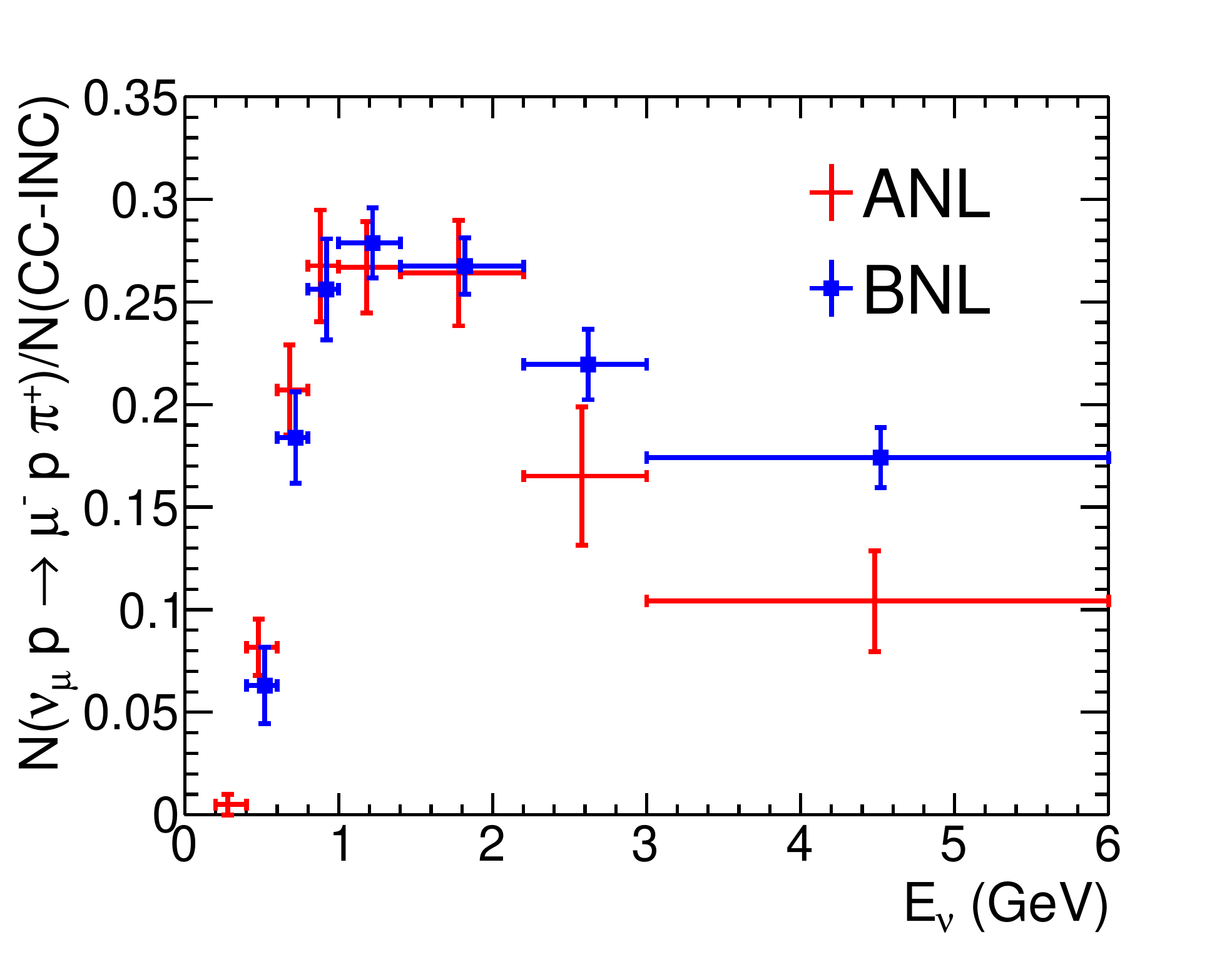}
  \caption{Ratio of \ccpi to CC-inclusive events as a function of $E_{\nu}$ for both ANL and BNL.}\label{fig:CC1pi_CCINC_ratio}
\end{figure}

We also find reasonable agreement in the ratio $\frac{\mathrm{CCQE}}{\mathrm{CC-INC}}$ as shown in Figure~\ref{fig:CCQE_CCINC_ratio}, and in the ratio $\frac{\mathrm{CC}1\pi^{+}}{\mathrm{CC-INC}}$ as shown in Figure~\ref{fig:CC1pi_CCINC_ratio}. However, CC-inclusive selections are more challenging than the exclusive channels CCQE and \ccpi. This is due to the high track multiplicity events which are included in CC-inclusive samples, and which have large uncertainties on their measuring efficiencies~\cite{BNL_Kitagaki_1986}. This can also be inferred from~\cite{ANL_Barish_1979} Table 1, where the corrected event rates for high track multiplicity events have large uncertainties. We also note that the correction factor for the BNL CC-inclusive dataset is based on our own estimate given in Section~\ref{sec:bnl}, as it was not published.

\FloatBarrier
\section{Conversion to cross-sections using known CCQE cross section}\label{sec:cross_sections}
As the CCQE cross-section on deuterium is relatively well understood, it is possible to produce \ccpi and CC-inclusive cross-section predictions by multiplying the cross-section ratios presented in the previous section by the CCQE cross-section. Effectively, this removes the ANL and BNL flux uncertainties, and replaces it with the theoretical uncertainty on the CCQE cross-section prediction, which is small (and has not been included in the plots presented here). The errors on our derived cross-sections are statistical only and may be larger than for the published ANL and BNL results, as the statistical error for two channels has been combined in quadrature.

\begin{figure}[htbp]
  \centering
  \includegraphics[width=0.45\textwidth]{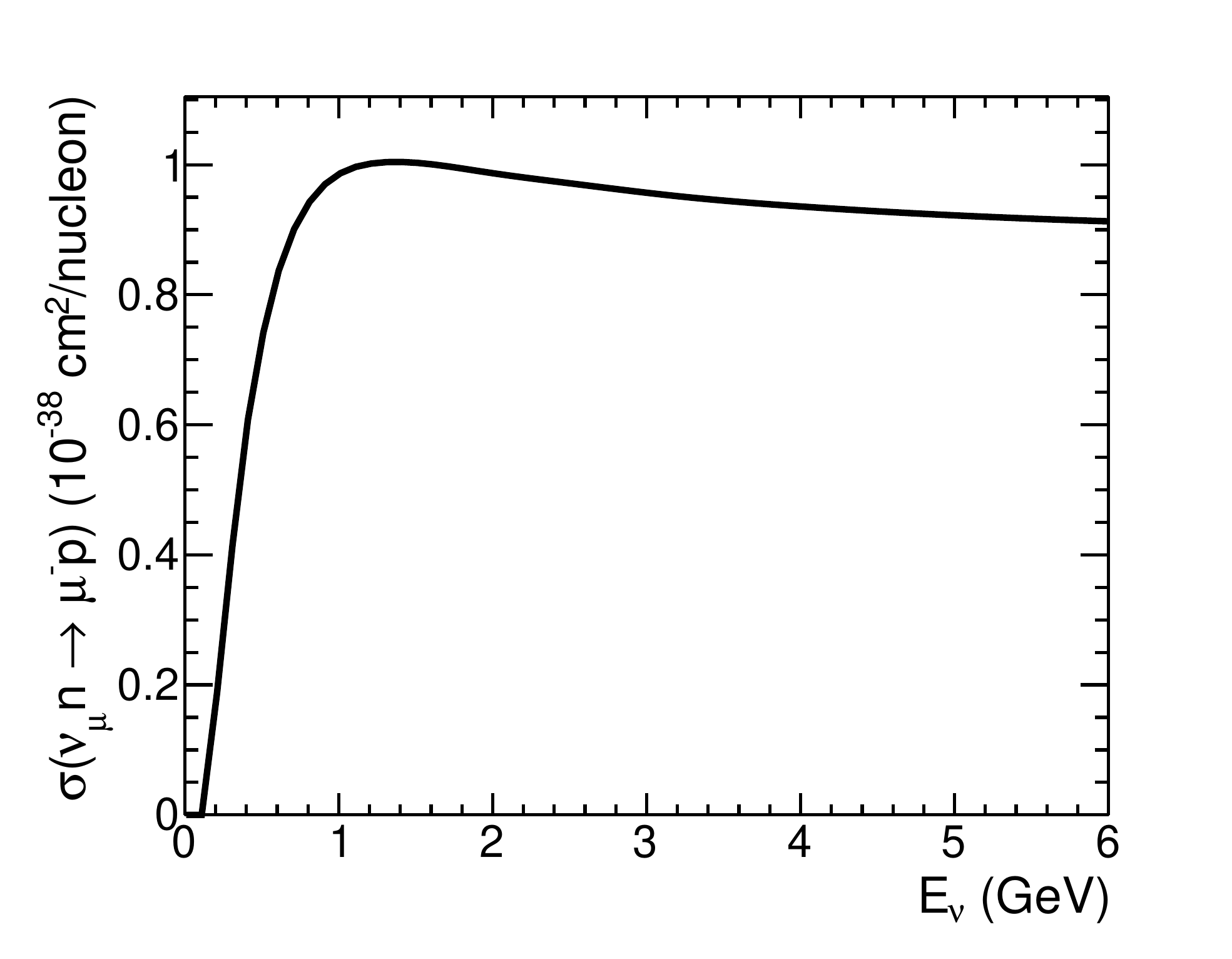}
  \caption{The default CCQE cross-section prediction for $\nu_{\mu}-\mathrm{D_{2}}$, taken from GENIE 2.8 using the default model parameters.}\label{fig:genie_spline}
\end{figure}

The $\nu_{\mu}-\mathrm{D_{2}}$ CCQE cross-section prediction we use is produced using GENIE 2.8~\cite{genieMC}, using the default Llewellyn~Smith~\cite{smith72} model parameters where the axial mass, $M_{A} = \unit[0.99]{GeV}$. This value is based on a fit to the shape of the $Q^2$ distribution of the ANL and BNL datasets~\cite{bba03}, which finds $M_{A} = 1.00 \pm \unit[0.02]{GeV}$. This fit is independent of the ANL and BNL flux normalizations, so our only assumption is that the Llewellyn~Smith model provides a reasonable description of CCQE neutrino--nucleon scattering. We note that another fit which includes the normalization and $E_{\nu}$ dependence for a large number of bubble chamber experiments finds a consistent result~\cite{kuzmin_2006} ($M_{A} = 0.96 \pm \unit[0.03]{GeV}$). The cross-section spline used in this analysis has been reproduced in Figure~\ref{fig:genie_spline}.

\begin{figure}[htbp]
  \centering
  \includegraphics[width=0.45\textwidth]{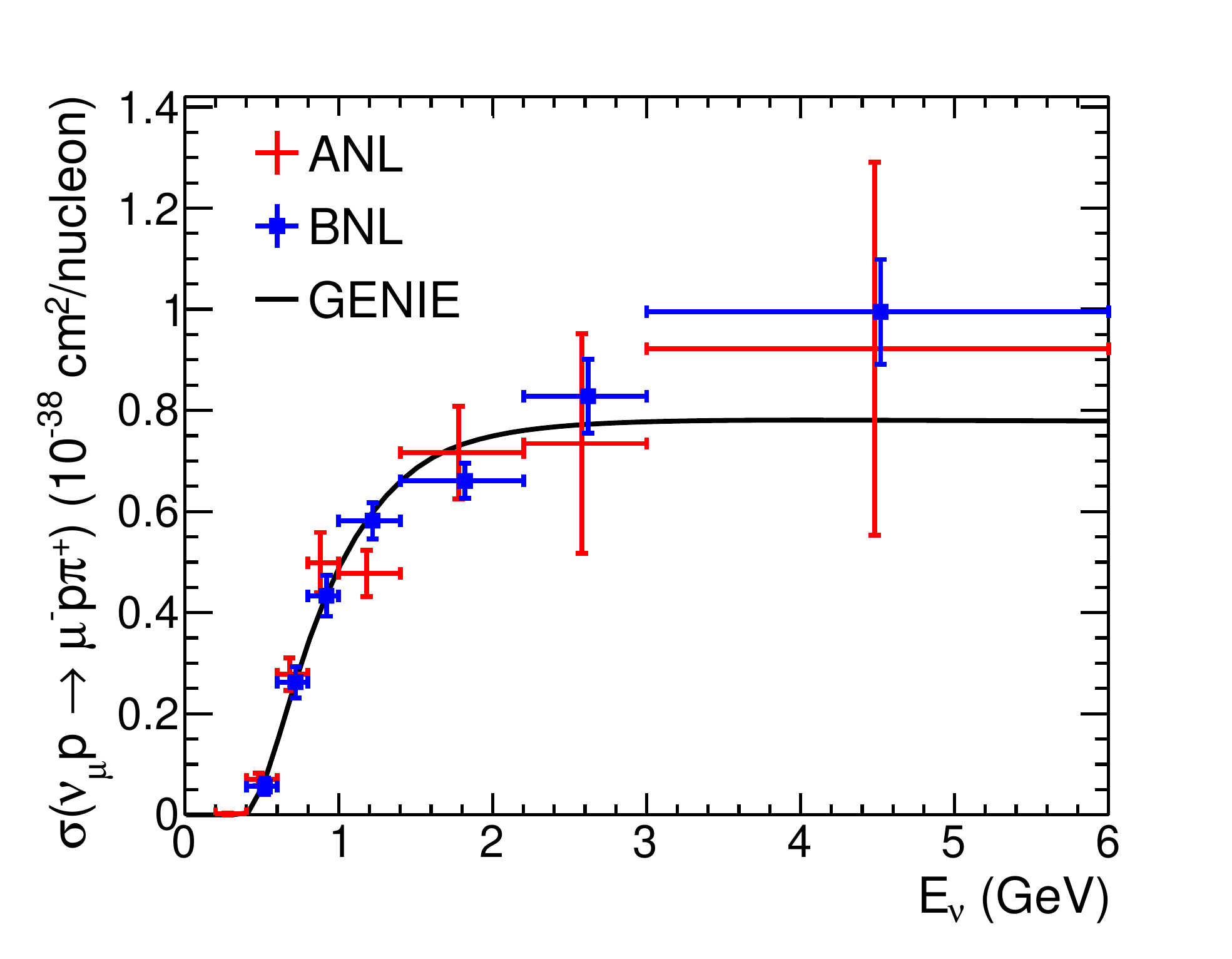}
  \caption{Comparison of the CC1$\pi^{+}$ cross-sections for both ANL and BNL, obtained by multiplying the ratio $\frac{\mathrm{CC1\pi^{+}}}{\mathrm{CCQE}}$ by the GENIE CCQE cross-section prediction for $\nu_{\mu}-\mathrm{D_{2}}$ interactions. The GENIE $\Delta^{++}$ cross-section prediction has been added for reference.}\label{fig:CC1pi_xsection}
\end{figure}
The $\mathrm{CC}1\pi^{+}$ cross-sections from both ANL and BNL, produced by multiplying the $\frac{\mathrm{CC}1\pi^{+}}{\mathrm{CCQE}}$ ratio by the GENIE CCQE cross-section, are shown in Figure~\ref{fig:CC1pi_xsection}. The GENIE $\Delta^{++}$ cross-section has been included for comparison, as this resonance makes the biggest contribution. However, higher order resonances also contribute to the measurements, particularly at high neutrino energies, so the measurements are expected to deviate from the GENIE predictions at high $E_{\nu}$. Note that there is no invariant mass cut on the distributions used to extract the \ccpi cross-sections produced in this work.
\begin{figure}[htbp]
  \centering
  \begin{subfigure}{0.45\textwidth}
    \includegraphics[width=\textwidth]{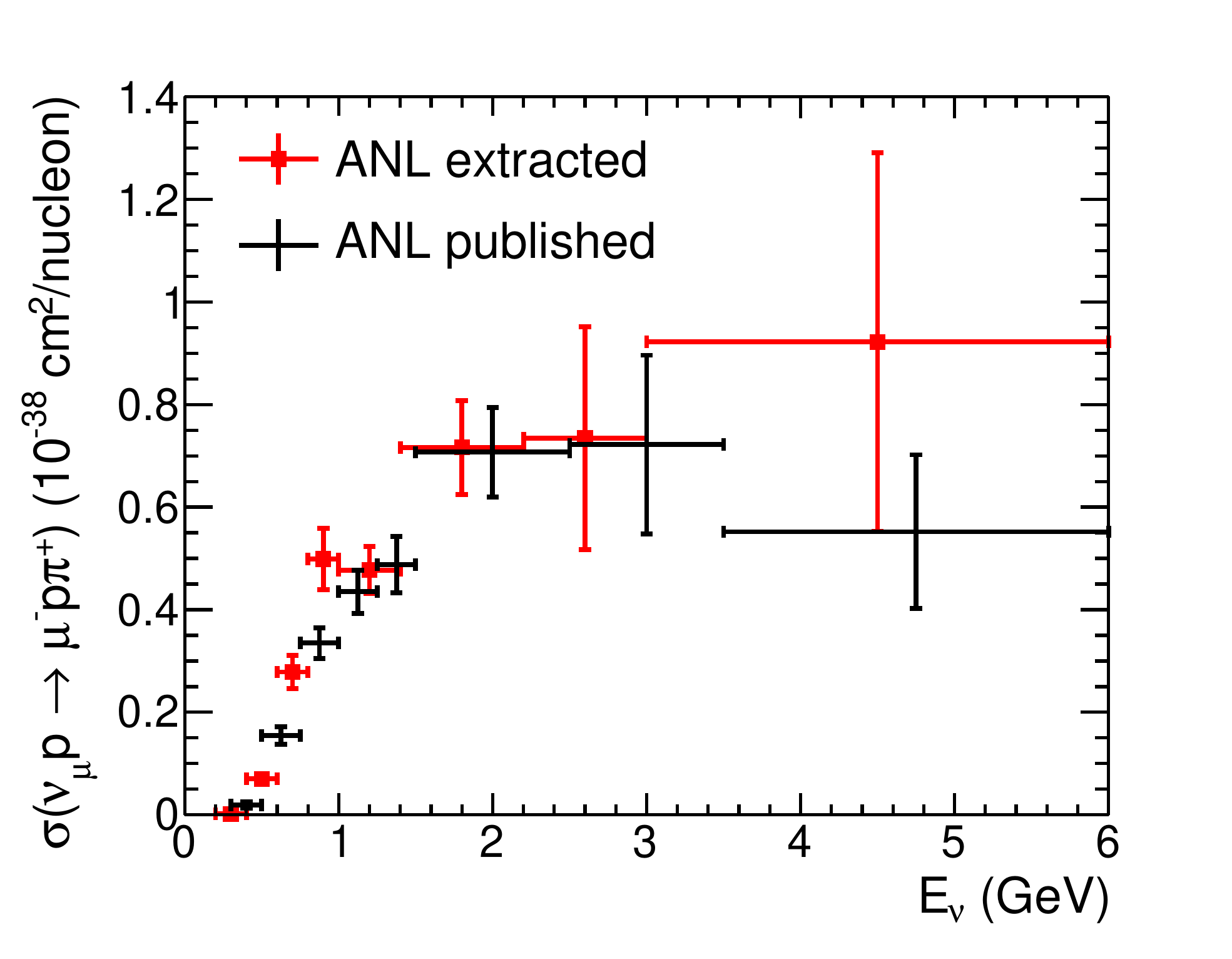}
    \caption{ANL}
    \label{subfig:CC1pi_xsection_ANL}
  \end{subfigure}
  \begin{subfigure}{0.45\textwidth}
    \includegraphics[width=\textwidth]{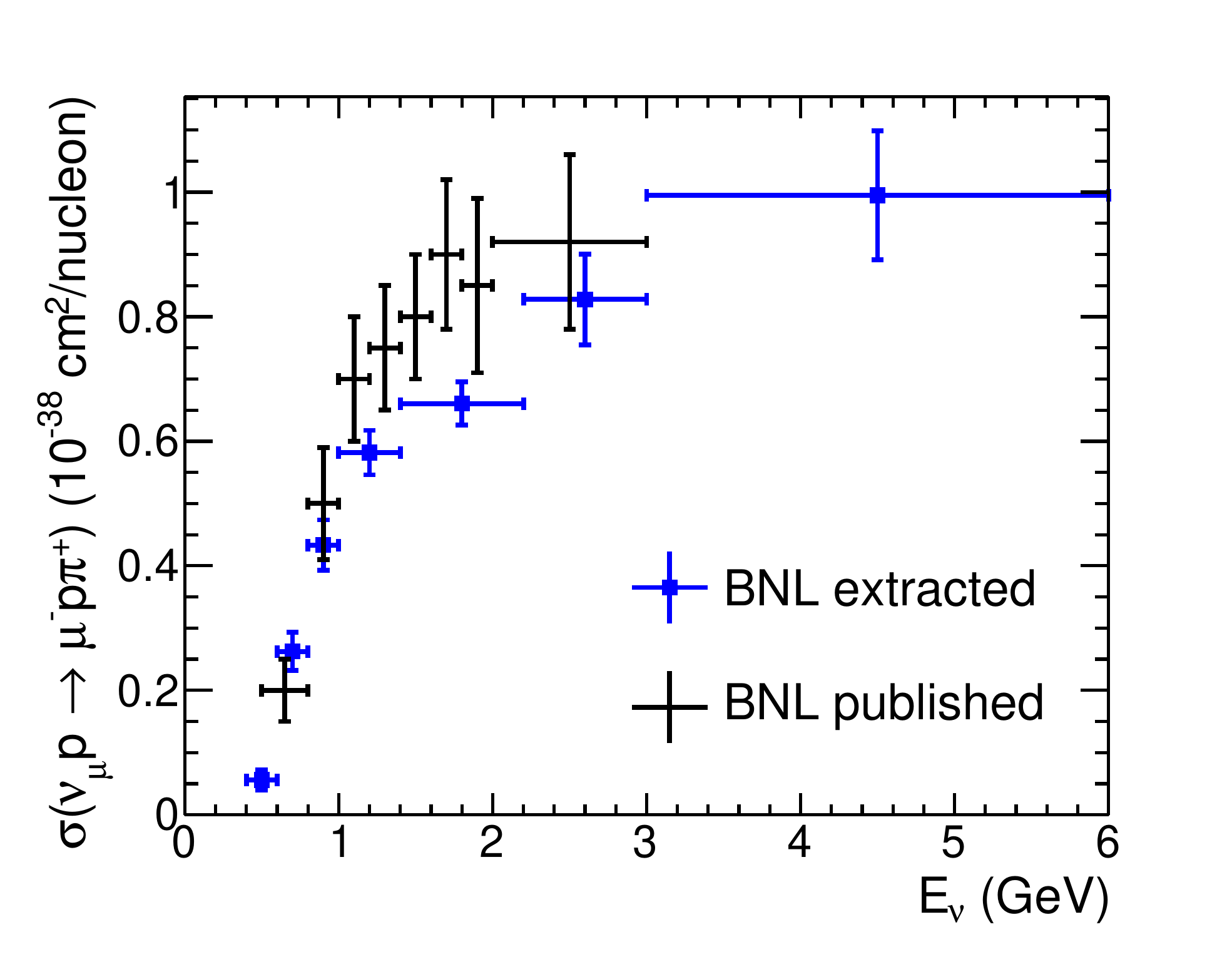}
    \caption{BNL}
    \label{subfig:CC1pi_xsection_BNL}
  \end{subfigure}
  \caption{Comparison of the CC1$\pi^{+}$ cross-sections for both ANL and BNL, compared with the published CC1$\pi^{+}$ cross-section from each experiment. Note that the published cross-section includes the flux normalization uncertainty.}\label{fig:CC1pi_xsection_comparisons}
\end{figure}

It is interesting to compare the extracted \ccpi cross-sections with those published by ANL and BNL, as shown in Figure~\ref{fig:CC1pi_xsection_comparisons}. In the neutrino energy range where ANL and BNL disagree most strongly, $1 \leq E_{\nu} \leq 2$ \unit{GeV}, the extracted BNL cross-section differs significantly from the published distribution (Figure~\ref{subfig:CC1pi_xsection_BNL}), whereas the extracted ANL results show reasonable agreement with the published ANL data (Figure~\ref{subfig:CC1pi_xsection_ANL}).

\begin{figure}[htbp]
  \begin{center}
  \includegraphics[width=0.45\textwidth]{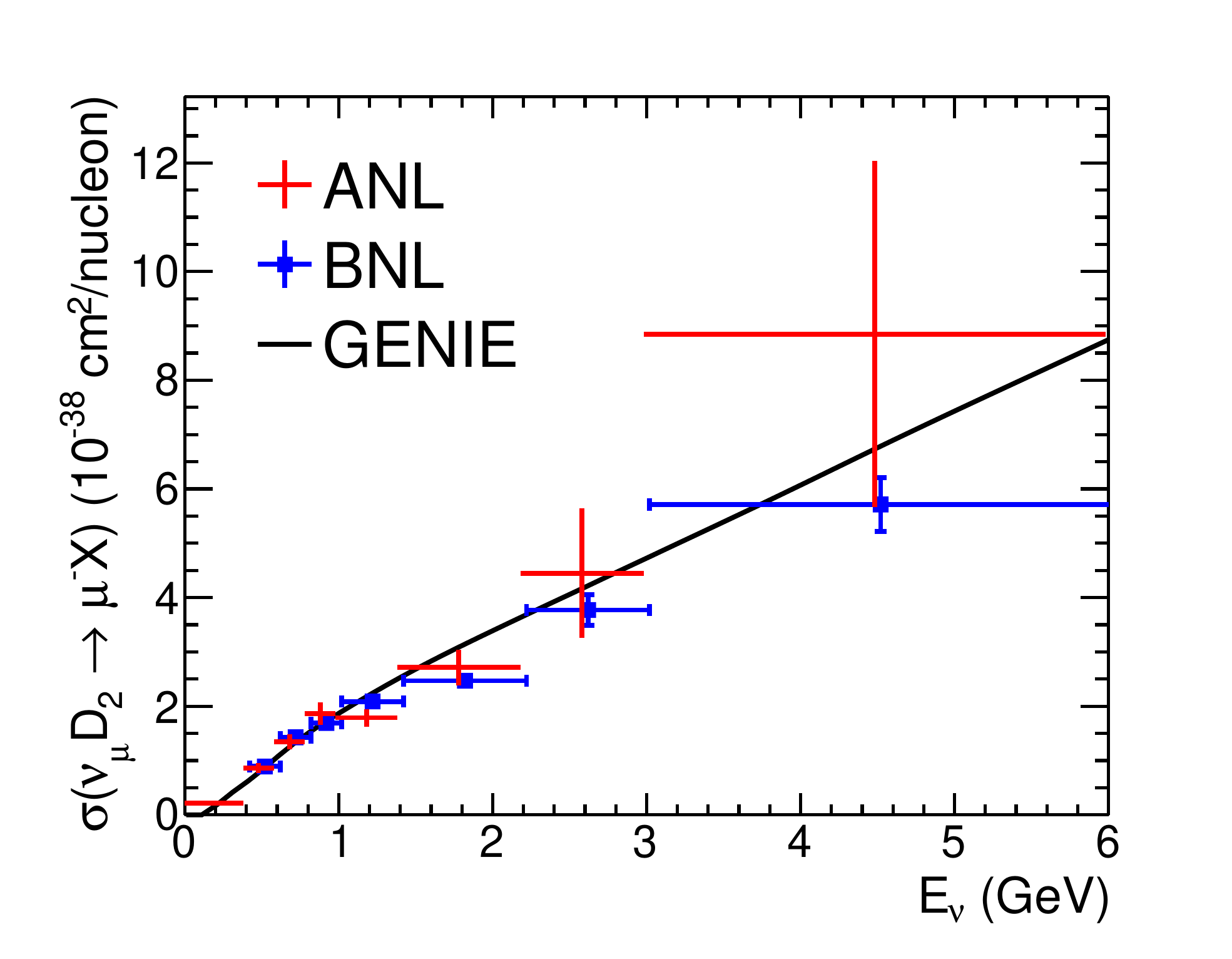}
  \end{center}
  \caption{Comparison of the \ccpi cross-sections for both ANL and BNL, obtained by multiplying the ratio $\frac{\mathrm{CC-inclusive}}{\mathrm{CCQE}}$ by the GENIE CCQE cross-section prediction for $\nu_{\mu}-\mathrm{D_{2}}$ interactions. The GENIE CC-inclusive cross-section prediction has been added for comparison, but was not used when producing the cross-sections.}\label{fig:CCINC_xsection}
\end{figure}

For completeness, Figure~\ref{fig:CCINC_xsection} shows the CC-inclusive cross-sections from both ANL and BNL, produced by multiplying the $\frac{\mathrm{CC-inclusive}}{\mathrm{CCQE}}$ ratio by the GENIE CCQE cross-section. However, we note again that the correction factor applied to the BNL CC-inclusive dataset was estimated as it was not explicitly given in a BNL publication.

\FloatBarrier
\section{Conclusions}\label{sec:conclusions}
In this work we have digitized and reanalysed ANL and BNL data for $\nu_{\mu}-\mathrm{D_{2}}$ scattering, and demonstrated that there is good agreement between ANL and BNL for the ratio $\sigma_{\ccpi}/\sigma_{\mathrm{CCQE}}$. This indicates that the outstanding ANL--BNL single pion production ``puzzle'' results from discrepancies in the flux predictions, which is in accordance with previous analyses of the same data~\cite{sobczyk_2009}\cite{sobczyk_2014}, which found that ANL and BNL agree within their published flux uncertainties. Using these ratios, we exploit the fact that the CCQE cross-section for interactions on deuterium is well understood to extract \ccpi cross-sections for both ANL and BNL. Although we only show statistical errors, the flux errors cancel, and the remaining normalization errors are small, and are likely to partially cancel when taking the ratio. Additional errors in the shape of the distributions from the energy resolution are likely to be small, and are unlikely to significantly distort the cross-section. Comparing our extracted results to the published ANL and BNL cross-sections, we found better agreement with ANL than BNL. However, we stress that both experiments gave large normalization uncertainties on their fluxes, so this is not indicative of a problem with the BNL results. The extracted cross-sections presented here resolve the longstanding ANL--BNL ``puzzle'', and should be used in future fits where this data is used to constrain the axial form factor for pion production on nucleons. The reduced error on this parameter will be of use to future neutrino oscillation measurements, and in interpreting the increasing body of single pion production data from nuclear targets~\cite{k2k-ccpi0, k2k-nc1pi0, sciboone_2010, mb-cc1pplus-2010, mb-ccpi0-2010, mb-ncpi0-2010, minerva_2014}, where nuclear effects have yet to be fully understood.

\begin{acknowledgements}
This work developed from studies performed by the T2K
experiment's neutrino interaction working group, who we thank for
their encouragement and comments, and discussions at the NuInt series
of conferences. We are grateful to Anthony Mann for helpful
comments on an earlier presentation of this work. This material is based upon work supported by the US
Department of Energy under Grant DE-SC0008475 and by the UK STFC as a PhD studentship.
\end{acknowledgements}

\bibliographystyle{apsrev4-1}
\bibliography{ANL_BNL_CC1pi}

\end{document}